# Owl Eyes: Spotting UI Display Issues via Visual Understanding

Zhe Liu[1,3], Chunyang Chen[4], Junjie Wang[1,2,3,*], Yuekai Huang[1,3], Jun Hu[1,3], Qing Wang[1,2,3,*]
[1]Laboratory for Internet Software Technologies, [2]State Key Laboratory of Computer Sciences,
Institute of Software Chinese Academy of Sciences, Beijing, China;
[3]University of Chinese Academy of Sciences, Beijing, China; [*]Corresponding author
[4]Monash University, Melbourne, Australia;
liuzhe181@mails.ucas.ac.cn,Chunyang.chen@monash.edu,junjie@iscas.ac.cn,wq@iscas.ac.cn

## ABSTRACT

Graphical User Interface (GUI) provides a visual bridge between a software application and end users, through which they can interact with each other. With the development of technology and aesthetics, the visual effects of the GUI are more and more attracting. However, such GUI complexity posts a great challenge to the GUI implementation. According to our pilot study of crowdtesting bug reports, display issues such as text overlap, blurred screen, missing image always occur during GUI rendering on different devices due to the software or hardware compatibility. They negatively influence the app usability, resulting in poor user experience. To detect these issues, we propose a novel approach, OwlEye, based on deep learning for modelling visual information of the GUI screenshot. Therefore, OwlEye can detect GUIs with display issues and also locate the detailed region of the issue in the given GUI for guiding developers to fix the bug. We manually construct a large-scale labelled dataset with 4,470 GUI screenshots with UI display issues and develop a heuristics-based data augmentation method for boosting the performance of our OwlEye. The evaluation demonstrates that our OwlEye can achieve 85% precision and 84% recall in detecting UI display issues, and 90% accuracy in localizing these issues. We also evaluate OwlEye with popular Android apps on Google Play and F-droid, and successfully uncover 57 previously-undetected UI display issues with 26 of them being confirmed or fixed so far.

## 1 INTRODUCTION

Graphic User Interface (GUI, also short for UI) is ubiquitous in almost all modern desktop software and mobile applications. It provides a visual bridge between a software application and end users through which they can interact with each other. Designing a UI requires proper user interaction, information architecture and visual effects of the UI. A good GUI design makes an application easy, practical and efficient to use, which significantly affects the success of the application and the loyalty of its users [27].

However, more and more fancy visual effects in GUI design such as intensive media embedding, animation, light and shadows post a great challenge for developers in the implementation. Consequently, many display issues[1] such as *text overlap, missing image, blurred screen* as seen in Figure 1 always occur during the UI display process especially on different mobile devices.

---
[1]we call these bugs as UI display issues, and will interchangably use *bug* and *issue* in this paper.

Most of those UI display issues are caused by different system settings in different devices, especially for Android, as there are more than 10 major versions of Android OS running on 24,000+ distinct device models with different screen resolutions [64]. Although the software can still run along with these bugs, they negatively influence the fluent usage with the app, resulting in the significantly bad user experience and corresponding loss of users. Therefore, this study is targeting at detecting those UI display issues.

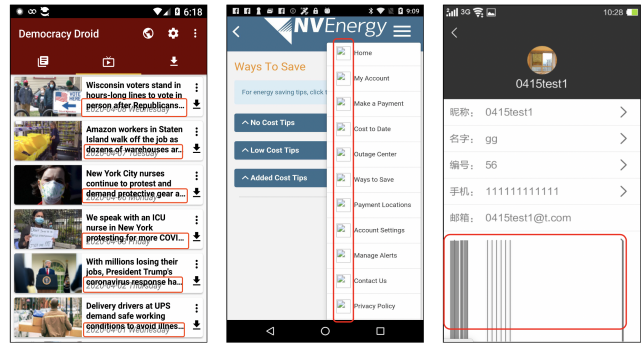

Figure 1: Examples of UI display issues

To ensure the correctness of UI displaying, companies have to recruit many testers for app GUI testing or leverage the crowdtesting. Although human testers can spot these UI display issues, there are still two problems with such mechanism. First, it requires significant human effort as testers have to manually explore tens of pages by different interactive ways and also need to check the UI display on different OS versions and devices with different resolution or screen size. Second, some errors in the GUI display, especially relatively minor ones such as *text overlap, component occlusion*, are difficult to spot manually. To overcome those issues, some app development teams adopt the Rapid Application Development (RAD) [43], which focuses on developing applications rapidly through frequent iterations and continuous feedback. They utilize users' feedback to reveal UI display issues, but it is a reactive way for bug fixing which may have already hurt users of the app, resulting in the loss of market shares.

In comparison with obtaining feedback from users for reactive app UI assurance, we need a more proactive mechanism which could check the UI display before the app release, automatically spot the potential issues in the GUI, and remind the developers to fix issues if any. There are many research works on automated GUI testing [3, 17, 20, 24, 39, 42, 44, 45, 48, 49, 60] by dynamically exploring different pages with random actions (e.g., clicking, scrolling, filling in the text) until triggering the crash bugs or explicit exceptions. Some practical automated testing tools like Monkey [21, 65],

Dynodroid [40] are also widely used in industry. However, these automated tools can only spot critical crash bugs, rather than UI display issues which cannot be captured by the system. In this work, we aim at detecting the UI display issues with the screenshots generated during automatic testing by visual understanding.

To understand the common UI rendering issues, we first carry out a pilot study on 10,330 non-duplicate screenshots from 562 mobile application crowdtesting tasks to observe display issues in these screenshots. Results show that a non-negligible portion (43.2%) of screenshots are of display issues which can seriously impact the user experience, and degrade the reputation of the applications. Besides, we also examine 1,432 screenshots randomly-chosen from the commonly-used Rico dataset [19], and find 1.2% screenshots having UI display issues. The common categories of UI display issues include *component occlusion, text overlap, missing image, null value* and *blurred screen*. Considering its popularity and lack of support in current practice of automatic UI testing, it would be valuable for classifying the screenshots with UI display issues from the plenty of screenshots generated during UI testing.

Inspired by the fact that display bugs can be easily spotted by human eyes, we propose an approach, OwlEye[2] to model the visual information by deep learning to automatically detect and localize UI display issues. Our OwlEye builds on the Convolutional Neural Network (CNN) to identify the screenshots with UI display issues, and utilizes Gradient weighted Class Activation Mapping (Grad-CAM) to localize the regions with UI display issues in the screenshots for guiding developers to fix the bug. To overcome the lack of labeled data for training our model, we develop a heuristics-based data augmentation method to generate screenshots with UI display issues from bug-free UI images. We then integrate OwlEye with DroidBot [36] which can dynamically explore different pages of the mobile apps as a fully automatic tool from collecting the screenshots to detect and localize UI display issues. Note that one strength of our approach over conventional program analysis is that it can be applied to any platform including Android, iOS, and it takes the screenshot as the input which is easy to be obtained in real-world practice.

To evaluate the effectiveness of our OwlEye, we carry out a large-scale experiment on 8,940 screenshots from crowdtesting and 15,640 augmented screenshots from 7,820 Android apps. Compared with 13 state-of-the-art baselines, our OwlEye can achieve more than 17% and 50% boost in recall and precision compared with the best baseline, resulting in 85% precision and 84% recall. As our OwlEye can also locate the detailed position of the bug in the UI, we carry out a user study to check its accuracy and the results demonstrate that 90% of bug locations are correct. Apart from the accuracy of our OwlEye, we also evaluate the usefulness of our OwlEye by applying it in detecting the UI display issues in the real-world apps from Google Play and F-Droid. Among 329 apps, we find that 57 of them are with UI display issues. We issued bug reports to the development team and 26 are confirmed and fixed by developers.

The contributions of this paper are as follows:

- This is the first work to conduct a systematical investigation of UI display issues in real-world mobile apps. We develop a first large-scale dataset of app UIs with that kind of bugs and release it for follow-up studies.
- Based on our pilot findings, we propose a novel approach OwlEye[3] with CNN-based model for detecting screenshots with UI display issues, and Grad CAM-based model for localizing the buggy region in the UI.
- We also propose a heuristics-based training data augmentation method which can automatically generate screenshots of UI display issues with bug-free UI images.

## 2 MOTIVATIONAL STUDY

To better understand the UI displaying issues in real-world practice, we carry out a pilot study to examine the prevalence of these issues. The pilot study also explores what kinds of UI display issues exist, so as to facilitate the design of our approach for detecting UIs with display issues.

### 2.1 Data Collection

Our experimental dataset is collected from one of the largest crowdtesting platforms[4] in which crowd workers are required to submit test reports after performing testing tasks [62, 63]. The dataset contains 562 Android mobile application crowdtesting tasks between January 2015 and September 2016. These apps belong to different categories such as news, entertainment, medical, etc. In each task, crowd workers submit hundreds of testing reports which describe how the testing is conducted and what happened during the test, as well as accompanied screenshots of the testing. The reason why we utilize this dataset is that it includes both the UI screenshots and the corresponding bug description which facilitates the searching and analysis of UI display issues. This dataset contains 10,330 unique GUI screenshots.

### 2.2 Categorizing UI Display Issues

Given those GUI screenshots, the first three authors individually check each of them manually with also its corresponding description in the bug report. Only GUI screenshots with the consensus from all three human markers are regarded as ones with display issues. A total of 4,470 GUI screenshots are determined with UI display issues, which accounts for 43.2% (4470/10330) in all screenshots. This result indicates that the UI display issues account for a non-negligible portion of mobile application bugs revealed during crowdtesting and should be paid careful attention for improving the software quality.

During the manually examination process, we notice that there are different types of UI issues, a categorization of these issues would facilitate the design and evaluation of related approach. Following the Card Sorting [59] method, we classify those UI issues into five categories including *component occlusion, text overlap, missing image, null value* and *blurred screen* with details as follows:

---

[2]Our approach is named as OwlEye as it is like owl's eyes to effectively spot UI display issues. And our model (nocturnal like owl) can complement with conventional automated GUI testing (diurnal like eagle) for ensuring the robustness of the UI.

[3]https://github.com/20200501/OwlEye for the dataset and source code of OwlEye, and the detailed experimental results of this paper.
[4]Baidu (baidu.com) is the largest Chinese search service provider. Its crowdsourcing test platform (test.baidu.com) is also the largest ones in China.

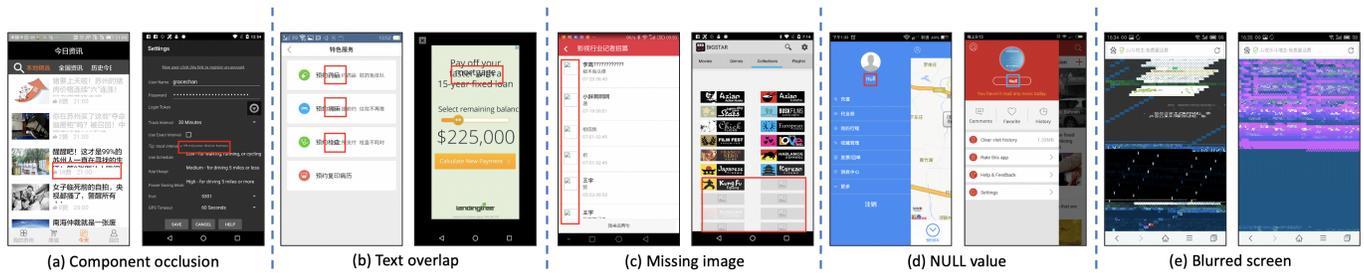

Figure 2: Examples of five categories of UI display issues

**Component occlusion (47%)**: As shown in Figure 2(a), the textual information or component is occluded by other components. It usually appears together with TextView or EditText. The main reasons are as follows: the improper setting of element's height, or the adaptive issues triggered when setting a larger-sized font.

**Text overlap (21%)**: As shown in Figure 2(b), two pieces of text are overlapped with each other. This might be caused by the adaptive issues among different device models, e.g., when using a larger-sized font in a device model with small screen might trigger this bug.

Note that, for text overlap category, two pieces of text are mixed together; while for component occlusion, one component covers part of the other component.

**Missing image (25%)**: As shown in Figure 2(c), in the icon position, the image is not showing as its design. The possible reasons are as follows: wrong image path or layout position, unsuccessful loading of the configuration file due to permissions, oversized image, network connection, code logic, or picture errors, etc.

**NULL value (6%)**: As shown in Figure 2(d), the right information is not displaying, instead *NULL* is showing in corresponding area. This category of bugs usually occurs with TextView. The main reasons are as follows: issues in parameter setting or database reading, and the length of text in TextView exceeding the threshold, etc.

**Blurred screen (1%)**: As shown in Figure 2(e), the screen is blurred. The reason for this bug might because the defects in hardware, or the exclusion of hardware acceleration for some CPU- or GPU- demanding functionalities.

To further validate the generality of our observations, we also manually check 1,432 screenshots from 200 random-chosen applications in Rico dataset[5] [19] , which is a commonly-used mobile application dataset with 66K UI screenshots of Android Applications and we will further introduce that dataset on Section 4. We find that 18 UIs from 16 apps (16/200 = 8.8% apps) are with UI display issues. Note that number is highly underestimated, as the collected UIs do not cover all pages of the applications, and the applications are not fully tested on different devices with different screen resolutions.

### 2.3 Why Visual Understanding in Detecting UI Display Issues

These findings confirm the severity of UI display issues, and motivate us to design approach for automatically detecting these GUI issues. One commonly-used practice for bug detection in mobile apps is program analysis, but it may not be suitable in this senario. To apply the program analysis, one need to instrument the target app, develop different rules for different types of UI display issues, rewrite the code for different platforms (e.g., iOS, Android), and customize their code to be compatible on different mobile devices (e.g., Samsung, Huawei, etc) with different screen resolution, which is extremely effort-consuming. Specifically, it is not trivial to enumerate all display issues and develop corresponding rules for detection.

Taken in this sense, it is worthwhile developing a new efficient and general method for detecting UI display issues. Inspired by the fact that these display issues can be spotted by human eyes, we propose to identify these buggy screenshots with visual understanding technique which imitates the human visual system. As the UI screenshots are easy to fetch (either manually or automatically) and exert no significant difference across the apps from different platforms or devices, our image-based approach are more flexible and easy to deploy.

## 3 ISSUES DETECTION AND LOCALIZATION APPROACH

This paper proposes OwlEye to automatically detect and localize UI display issues in the screenshots of the application under test, as shown in Figure 3. Given one UI screenshot, our CNN-based model can first classify if it relates with any display issues via the visual understanding. Once the issue is confirmed, our model can further localize the detailed issue position on the UI screenshot by Grad CAM-based model for guiding developers to fix the bug.

### 3.1 CNN-based UI Display Issues Detection

As the UI display issues can only be spotted via the visual information, we adopt the convolutional neural network (CNN) [31, 34], which has proven to be effective in image classification and recognition in computer vision [25, 58, 61]. Figure 4 shows the structure of our classification model which links the convolutional layers, batch normalization layers, pooling layers, and fully-connected layers.

Given the input UI screenshot, we convert it into a certain image size with fixed width and height as $w \times h$. Convolutional layer's parameters consist of a set of learnable filters. The purpose of the convolutional operation is to extract the different characteristics of the input (i.e., feature extraction). After convolutional layer, the screenshots will be abstracted as feature graph.

In order to improve the performance and stability of CNN, we add Batch Normalization (BN) [26] after convolutional layer, and standardize the input layer by adjusting and scaling activation. In a neural network, batch normalization is implemented through a normalization step that fixes the mean and variance of each layer's inputs. In detail, the steps for batch normalization are shown below:

$$y = \frac{f - mean(f)}{\sqrt{var(f) + \epsilon}} \tag{1}$$

---
[5]http://interactionmining.org/rico#quick-downloads

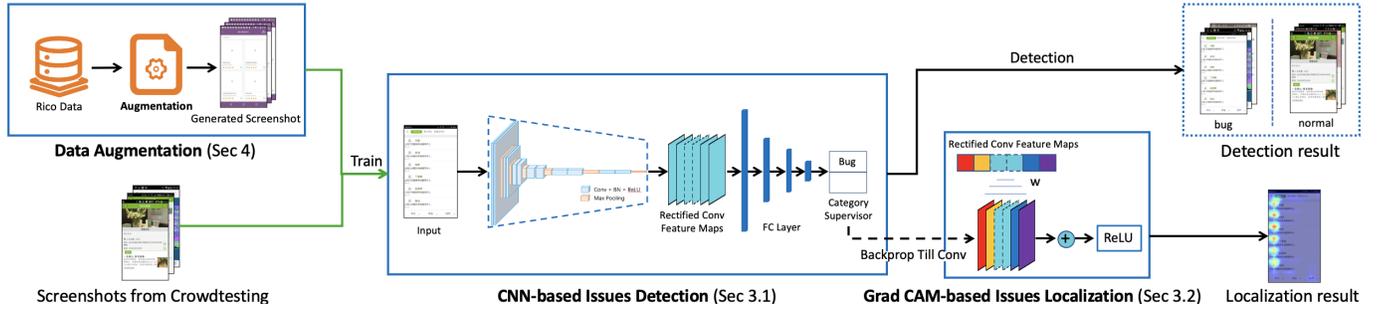

Figure 3: Overview of `OwlEye`

Considering a batch training, we input feature as $f$, then calculate the mean($mean()$) and variance($var()$) of $f$, $\epsilon$ is added in the denominator for numerical stability and is an arbitrarily small constant.

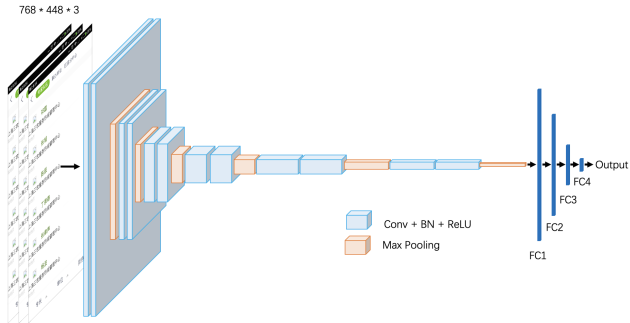

Figure 4: The architecture of `CNN`

After the BN layer, the Rectified Linear Unit (ReLU) is added as the activation function of the network. It increases the nonlinear properties of the decision function and of the overall network without affecting the receptive fields. ReLU performs a threshold operation to each element of the input, where any value less than zero is set to zero.

The BN layer is then followed by the pooling layer, which is to further pick up larger-scale detail than just edges and curves by further distilling the features. The pooling function uses the total statistical characteristics of the adjacent output of a certain location of the inputted image to replace the output of the network at that location, and combines the output of one layer of neuron cluster into a single neuron in the next layer to reduce the size of data. Max pooling uses the maximum value from each of a cluster of neurons at the prior layer. In a CNN's pooling layers, feature maps are divided into rectangular sub-regions, and the features in each rectangle are independently down-sampled to a single value, commonly by taking their average or maximum value. In addition to reducing the sizes of feature maps, the pooling operation grants a degree of translational invariance to the features contained therein.

The last several layers are fully connected neural networks (FC) which compile the data extracted by previous layers to form the final output. All inputs from one of these layers are connected to every activation unit of the next layer. The multiple fully connected relationships increase the possibility of learning a complex function. The fully connected layers further encode all features of the UI screenshot into a $K$-dimensional vector. Finally, the detection results are obtained through softmax [7].

$$P(y = b|f) = \frac{e^{f^T w_b}}{\sum_{k=1}^{K} e^{f^T w_k}} \tag{2}$$

where the $K$-dimensional vector are normalized into a probability distribution with $K$ probabilities, which is proportional to the index of the input number. The input $f$ is the feature, and $P(y = b|f)$ is the predicted probability of $f$ belonging to category $b$ (bug), which is similar to the result of the previous layer.

### 3.2 Grad CAM-based UI Display Issues Localization

Although our classification model can check if the given UI screenshot is of display issues, some UI display issues may still be too small to spot in a large UI screenshot. Therefore, besides the classification model, we adopt the feature visualization method to locate the detailed position of the issues for guiding developers to fix the bug. This can also help us evaluate whether the feature extracted by our CNN model is accurate or not. We apply Grad-CAM model for the localization of UI display issues. Gradient weighted Class Activation Mapping (Grad-CAM) is a technique for visualizing the regions of input that are "important" for predictions on CNN-based models [55] . The final convolutional layer of CNN model contains the spatial and semantic information, and this technique uses the class-specific gradient information flowing into the final convolutional layer to produce a localization map of the important regions in the inputted image. The flow of Grad-CAM is shown in Figure 5. First, a screenshot with UI display issue is input into the trained CNN model, and the category supervisor to which the image belongs is set to 1, while the rest is 0.

Then the information is propagated back to the convolutional feature map of interest to obtain the Grad-CAM positioning. Suppose that the judgment category is $b$ (Bug), the calculation method of the score gradient of $b$ is $\frac{\partial output^b}{\partial A_{ij}^K}$, where $output^b$ is the output of category $b$ before softmax. Through the feedback of global average pooling of the gradient, the weight $\alpha_k^b$ of the importance of neurons is obtained. This weight captures the importance of the feature map $K$ of the target class $b$. By performing the weighted combination of the forward activation graph, we can obtain the class-discriminative localization map $L_{Grad-CAM}^b$.

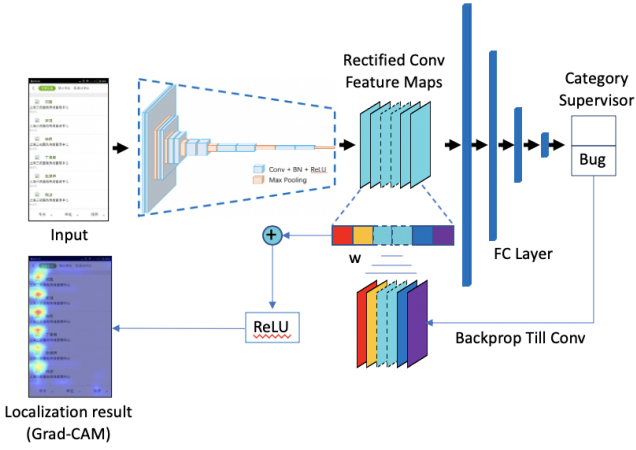

Figure 5: The architecture of Grad-CAM

$$\alpha_k^b = \frac{1}{Z} \sum_i \sum_j \frac{\partial output^b}{\partial A_{ij}^K} \quad (3)$$

$$L_{Grad-CAM}^b = ReLU(\sum_k \alpha_k^c A^k) \quad (4)$$

Finally, the point multiplication with the back propagation can obtain the Grad-CAM as the result of UI display issues localization.

### 3.3 Implementation

Our CNN model is composed of 12 Convolutional layers with batch normalization, 6 pooling layers and 4 full connection layers for classifying UI screenshot with display issues. The size of convolutional kernel in convolutional layer is 3 * 3. We set up the number of convolutional kernels as 16 for convolutional layer 1-4, 32 for convolutional layer 5-6, 64 for convolutional layer 7-8, and 128 for convolutional layer 9-12. The momentum in BN layer is set as 0.1. For the pooling layers, we use the most common-used max-pooling setting [57], i.e., pooling units of size 2 × 2 applied with a stride [58]. We set the number of neurons in each of the fully connected layers as 4096, 1024, 128 and 2 respectively. For data preprocessing, we rotate some UI of the horizontal screens to vertical, and resize the screens to 768 * 448. We implement our model based on the PyTorch[6] framework.

## 4 HEURISTIC-BASED DATA AUGMENTATION

Training an effective CNN model for visual understanding requires a large amount of input data. For example, RESNET [25] model uses 128 million images from ImageNet as training dataset for image classification task. Similarly, training our proposed CNN for UI display issues detection requires abundant of screenshots with UI display issues. However, there is so far no such type of open dataset, and collecting the related buggy screenshots is quite time- and effort-consuming. Therefore, we develop a heuristic-based data augmentation method for generating UI screenshots with display issues from bug-free UI images.

The data augmentation is based on the Rico [19] dataset which contains more than 66K unique screenshots from 9.3K Android applications, as well as their accompanied JSON file (i.e., detailed run-time view hierarchy of the screenshot). According to our observation on Section 2, most UI screenshots in this dataset are of no dispaly issues.

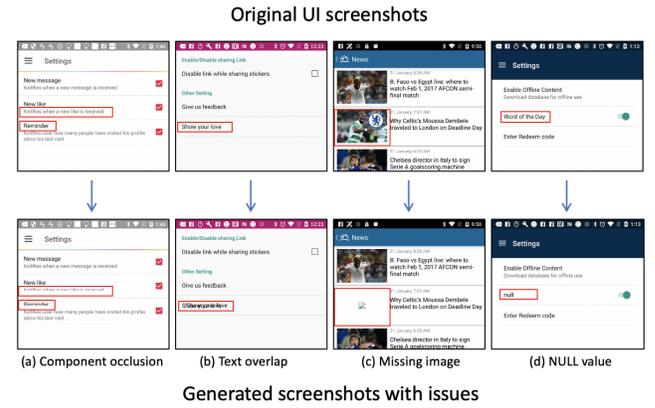

Figure 6: Examples of data augmentation

Algorithm 1 presents the heuristic-based data augmentation algorithm. With the input screenshot and its associated JSON file, the algorithm first locates all the TextView and ImageView, then randomly chooses a TextView or ImageView depending on the augmented category. Based on the coordinates and size of the TextView/ImageView, the algorithm then makes its copy and adjusts its location or size following specific rules to generate the screenshot with corresponding UI display issues. Figure 6 demonstrates the illustrative examples of the augmented screenshots with UI display issues.

Note that, among the five categories of UI display issues, the category of *blurred screen* is difficult to be generated following the above idea. Besides, preliminary results reveal the proposed approach can detect this category of issues with relatively high accuracy. Hence, we leave this category for future work, and in this study we obtain this category of screenshots online by searching '*blurred screen*'. We then present the detailed augmentation rules of the four categories.

**Augmentation for Component Occlusion Bug**: When this category of bug occurs, the textual information or component is occluded by other components. Therefore, we first generate a color block which shares the same background color as the original TextView but with a smaller height, then put it to cover part of the TextView randomly.

**Augmentation for Text Overlap Bug**: The textual contents are overlapped with each other, when this category of bug occurs. To augment this category of screenshots, we generate a piece of text with the same content as the original TextView, and offset it slightly.

---
[6]https://pytorch.org

---

**Algorithm 1:** Heuristic-based data augmentation

**Input:** *scr*: screenshot without bugs;
*json*: associated JSON file;
*category*: category of generated UI display issue;
*icon*: pre-prepared image icon;
**Output:** *augscr*: augmented screenshot with *category* bug;

1  Traverse *json* file to obtain all TextView and ImageView;
2  **if** *category* == 'missing image' **then**
3      Randomly choose an ImageView;
4  **else**
5      Randomly choose a TextView;
6  Obtain the coordinates of TextView/ImageView $(x_1,y_1),(x_2,y_2)$; //coordinate of upper left and lower right
7  Calculate the width and height of TextView/ImageView ($w$, $h$) based on the coordinates;
8  Obtain the text content of TextView (*text*);
9  Obtain the background color of TextView/ImageView (*bg*);
10 **if** *category* == 'component occlusion' **then**
11     $rand \leftarrow$ random.uniform(-1,1);
12     *image*.new(($w, h \times |rand|$),*bg*);
13     **if** $rand \geq 0$ **then**
14         //Occlude the upper part of component
            $augscr \leftarrow scr$.paste(*image*, $(x_1,y_1)$);
15     **else**
16         //Occlude the lower part of component
            $augscr \leftarrow scr$.paste(*image*, $(x_1,y_2 + (h \times rand))$);
17 **if** *category* == 'text overlap' **then**
18     $xrand \leftarrow$ random.uniform($-0.5 \times w, 0.5 \times w$);
19     $augscr \leftarrow scr$.write($[x_2 - xrand, y_1]$,*text*);
20 **if** *category* == 'missing image' **then**
21     *image*.new(($w, h$),*bg*);
22     $scr$.paste(*image*,$(x_1,y_1)$);
23     $augscr \leftarrow scr$.paste(*icon*,$(x_1 + 0.5 \times w, y_1 + 0.5 \times h)$);
24 **if** *category* == 'null value' **then**
25     *image*.new(($w, h$),*bg*);
26     $scr$.paste(*image*, $(x_1,y_1)$);
27     $augscr \leftarrow scr$.write($[x_1,y_1]$,"null");
28 **return** *augscr*;

---

**Augmentation for Missing Image Bug**: We notice that when this category of bug occurs, an image icon would show up to indicate that the area supposes to be an image. To augment this category of screenshots, we first download some frequently-used image icons online, then cover the original image displaying area with one random-chosen image icon and set its background color as the color of its original image.

**Augmentation for NULL Value Bug**: When this category of bug occurs, *NULL* is displayed in the area where supposes to be a piece of text. We generate this category of screenshots by covering the original TextView using a color block which shares the same background color and with *NULL* on it.

Note that, both *component occlusion* and *text overlap* involves covering a TextView, the difference is that the former one utilizes a color block to cover the TextView so that it looks like a component blocks the text, while the latter one employs a piece of text to cover the TextView to make it look like the two pieces of text are overlapped with each other. Another note is that, based on our observation on the screenshots with UI display issues in Section 2, when conducting the augmentation, the TextView is covered in the vertical direction in *component occlusion*, while it is covered in the horizontal direction in *text overlap*.

## 5 EXPERIMENT DESIGN

### 5.1 Research Questions

- **RQ1: (Issues Detection Performance)** How effective of our proposed `OwlEye` in detecting UI display issues?

  For RQ1, we first present some general views of our proposed approach for UI display issues detection and the comparison with commonly-used baseline approaches (details are in Section 5.3). We also present the performance comparison among the variations of model configuration (e.g., the number of convolutional layers) to further demonstrate its effectiveness. Besides, we also evaluate the contribution of data argumentation by comparing the performance with and without the argumented training data.

- **RQ2: (Issues Localization Performance)** How effective of our proposed `OwlEye` in localizing UI display issues?

  For evaluating the performance of issues localization, we conduct a user study to check its accuracy.

- **RQ3: (Usefulness Evaluation)** How does our proposed `OwlEye` work in real-world situations?

  For RQ3, we integrate `OwlEye` with DroidBot as a fully automatic tool to collect the screenshots and detect UI display issues, and then issue the detected bugs to the development team.

### 5.2 Experimental Setup

The experimental dataset comes from two sources. The first is the screenshots from crowdtesting, which contains 4,470 non-duplicate screenshots with UI display issues and equal number of bug-free non-duplicate screenshots (see details in Section 2).

The second is the screenshots generated with the data augmentation method in Section 4. In detail, we randomly download one screenshot from each of the random-chosen 10,000 applications in Rico dataset, and each screenshot would be utilized once for the data augmentation. In order to make the the training data balanced across categories, we use 10% screenshots for augmenting the *component occlusion* category, while use 30% screenshots for data augmentation of each of the other three categories.

For the augmented 10,000 screenshots with UI display issues, we first extract their features with ORB feature extraction algorithm [54], rank them randomly, compute the cosine similarity between a specific screenshot and each of its previous ones, and remove it when a similarity value above 0.8 is observed. In this way, 7,800 screenshots with UI display issues are remained and added into the experimental dataset. To make the data balanced, we then randomly download the screenshots from Rico and remove the similar ones, and a total of 7,800 bug-free screenshots are collected for experiment. For the category *blurred screen*, we randomly download 20 screenshots with this issue online, and randomly choose equal number of bug-free screenshots from Rico. Note that, the cosine similarity between each pair of these 40 screenshots is also below 0.8.

Table 1: The number of 5 categories of buggy screenshots

| Category | Train | | Test | Val |
|---|---|---|---|---|
| | Crowd-data | Aug-data | | |
| Component occlusion | 1745 | 986 | 226 | 131 |
| Text overlap | 706 | 2300 | 145 | 87 |
| Missing image | 569 | 2326 | 326 | 231 |
| NULL value | 130 | 2188 | 73 | 49 |
| Blurred screen | 20 | 20 | 30 | 2 |
| Overall | 3170 | 7820 | 800 | 500 |

In order to simulate the real-world application of our proposed approach, we setup the experiment as follows.

For the 8,940 screenshots screenshots from 562 crowdtesting apps, we utilize the 1,600 screenshots (800 with UI display issues and 800 without) from 162 apps as testing set to evaluate the performance of OwlEye, and employ another 1,000 screenshots (half of them with UI display issues) from another 50 apps as validation set to estimate how well the model has been trained and further tune the parameters. The 6,340 screenshots from the remaining 350 apps is utilized as training set. Besides, all the 15,640 screenshots (half of them with UI display issues) generated with data augmentation is added to the training set to boost the detection performance. Table 1 presents the distribution of screenshots in terms of different categories. The model is trained in a NVIDIA GeForce RTX 2060 GPU (16G memory) with 100 epochs for about 8 hours.

### 5.3 Baselines

In order to further demonstrate the advantage of OwlEye, we compare it with 13 baselines utilizing both machine learning and deep learning techniques. The 12 machine learning approaches first extract visual features from the screenshots, and employ machine learner for the classification. The deep learning approach utilizes artificial neural network directly on the screenshots for classification. We first present the three types of feature extraction method used in machine learning approaches.

**SIFT** [37]: Scale invariant feature transform (SIFT) is a common feature extraction method to detect and describe local features in an image. It can extract the interesting points on the object to generate the feature description of the object, which is invariant to uniform scaling, orientation, and illumination changes.

**SURF** [4]: Speed up robot features (SURF) is an improvement of *SIFT*. SURF uses an integer approximation of the determinant of Hessian blob detector, which can be computed with 3 integer operations using a precomputed integral image.

**ORB** [54]: Oriented fast and rotated brief (ORB) is a fast feature point extraction and description algorithm. Based on the rapid binary descriptor ORB of brief, it has rotation invariance and anti noise ability.

With these features, we apply four commonly-used machine learning approaches, i.e., Support Vector Machine (SVM) [30], K-Nearest Neighbor (KNN) [6], Naive Bayes (NB) [30] and Random Forests (RF) [8], for classifying the screenshots with UI display issues. The combination of three types of image features and four learning algorithms generates a total of 12 baselines.

We also experiment with Multilayer Perceptron (MLP) directly on the screenshots to better demonstrate the superiority of our proposed approach. In detail, MLP is a feedforward artificial neural network [23, 33]. The network structure is divided into input layer, hidden layer and output layer. Each node is a neuron that uses a nonlinear activation function, e.g., corrected linear unit (ReLU). It is trained by changing the connection weight according to the output error compared with the ground truth. We used eight layers of neural network, and we set the number of neurons in each layer to 190, 190, 128, 128, 64, 64, 32 and 2, respectively.

### 5.4 Evaluation Metrics

In order to evaluate the issues detection performance of our proposed approach, we employ three evaluation metrics, i.e., precision, recall, F1-Score, which are commonly-used in image classification and pattern recognition [38, 41]. For all the metrics, higher value leads to better performance.

Precision is the proportion of screenshots that are correctly predicted as having UI display issues among all screenshots predicted as buggy:

$$precision = \frac{Screenshots\ correctly\ predicted\ as\ buggy}{All\ screenshots\ predicted\ as\ buggy} \quad (5)$$

Recall is the proportion of screenshots that are correctly predicted as buggy among all screenshots that really have UI display issues.

$$recall = \frac{Screenshots\ correctly\ predicted\ as\ buggy}{All\ screenshots\ really\ buggy} \quad (6)$$

F1-score (F-measure or F1) is the harmonic mean of precision and recall, which combines both of the two metrics above.

$$F1 - score = \frac{2 \times precision \times recall}{precision + recall} \quad (7)$$

In Section 6.2, we employ Kendall's W (Kendall's coefficient of concordance) [56] to assess the agreement of the user evaluated localization results among different practitioners. It is a commonly-used measurement for the level of agreement between multiple items of multiple raters. The closer the test outcome is to 1, the higher agreement among the evaluation results of the raters.

## 6 RESULTS AND ANALYSIS

### 6.1 Issues Detection Performance (RQ1)

We first present the issues detection performance of our proposed OwlEye, as well as the performance in terms of five categories of UI display issues in Table 2. With OwlEye, the precision is 0.85, indicating 85% (679/798) of the screenshots which are predicted as having UI display issues are truly buggy. The recall is 0.84, indicating 84% (679/800) buggy screenshots can be found with OwlEye.

**Table 2: Issues detection performance (RQ1)**

| Category | Precision | Recall | F1-score |
|---|---|---|---|
| **Overall** | **0.850** | **0.848** | **0.849** |
| Component occlusion | 0.859 | 0.814 | 0.836 |
| Text overlap | 0.818 | 0.806 | 0.812 |
| Missing image | 0.855 | 0.904 | 0.879 |
| NULL value | 0.855 | 0.808 | 0.830 |
| Blurred screen | 0.888 | 0.800 | 0.842 |

We then shift our focus to the bottom half of Table 2, i.e., the performance in terms of each category of UI display issues. All the five categories of UI display issues can be detected with a relative high precision and recall, i.e., mimimum precision and recall are 0.82 and 0.80 respectively. The category *missing image* can be detected with the highest F1-score, indicating both precision (0.86) and recall (0.90) achieve a relatively high value. This might because screenshots with *missing image* bugs have relatively fixed pattern and the buggy area is relatively large, i.e., the whole image icon as shown in Section 2. In comparison, the category *text overlap* is recognized with the lowest F1-score, e.g., 0.82 precision and 0.81 recall. This is due to the fact that the pattern of this category is more diversified, and the buggy region is much smaller, i.e., the overlapping area between two pieces of text accounts for a mere of 10% of the text component.

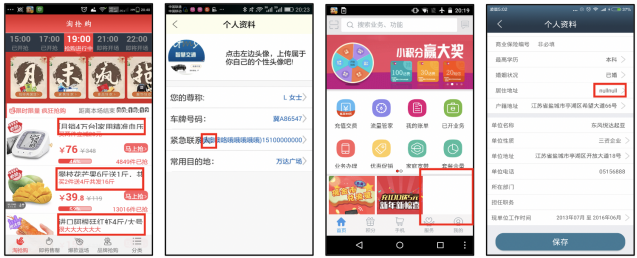

**Figure 7: Examples of bad case in issues detection (RQ1)**

We further analyze the screenshots which are wrongly predicted as bug-free, with examples in Figure 7. One common shared by these screenshots is that the buggy area is too tiny to be recognized even with human eye. Future work will focus more on improving the detection performance for these screenshots with attention mechanism and image magnification.

*6.1.1 Performance Comparison with Baselines.* Table 3 shows the performance comparison with the baselines. We can see that our proposed OwlEye is much better than the baselines, i.e., 17% higher in recall compared with the best baseline (MLP), and 50% higher in precision with the best baseline (ORB-NB). This further indicates the effectiveness of OwlEye. Besides, it also implies that OwlEye is especially good at hunting for the buggy screenshots from candidate ones, i.e., larger improvement in recall.

MLP achieves the highest recall among the baselines, indicating this deep learning approach is better at identifying the buggy screenshots, yet with a lower precision. The machine learning approaches with ORB feature achieve the highest F1-score (i.e., 0.63 by ORB-NB), indicating this kind of feature is more suitable for detecting UI display issues. This might because ORB algorithm is the state-of-the-art feature extraction algorithm, and has been proven to be an efficient alternative to SIFT or SURF [54].

**Table 3: Performance comparison with baselines (RQ1)**

| Method | Precision | Recall | F1-score |
|---|---|---|---|
| SIFT-SVM | 0.486 | 0.349 | 0.406 |
| SIFT-KNN | 0.510 | 0.492 | 0.501 |
| SIFT-NB | 0.584 | 0.411 | 0.482 |
| SIFT-RF | 0.458 | 0.458 | 0.432 |
| SURF-SVM | 0.561 | 0.512 | 0.535 |
| SURF-KNN | 0.522 | 0.526 | 0.524 |
| SURF-NB | 0.428 | 0.597 | 0.499 |
| SURF-RF | 0.513 | 0.524 | 0.519 |
| ORB-SVM | 0.551 | 0.514 | 0.532 |
| ORB-KNN | 0.525 | 0.522 | 0.523 |
| ORB-NB | 0.567 | 0.709 | 0.630 |
| ORB-RF | 0.520 | 0.528 | 0.524 |
| MLP | 0.537 | 0.727 | 0.618 |
| **OwlEye** | **0.850** | **0.848** | **0.849** |

*6.1.2 Performance Comparison among Model Configurations.* We also conduct experiments to compare the detection performance with different configurations of CNN model. Table 4 shows the performance of UI display issues detection in terms of different convolutional layers and with / without batch normalization (BN).

We can see that both the convolutional layers and the batch normalization can influence the issues detection performance. Generally speaking, when deepening the neural network, i.e., more convolutional layers, both precision (P) and recall (R) would increase. For example, the precision undergo 18% improvement when convolutional layers increase from 4 to 12 (with batch normalization), and the improvement of recall is 58% with same configuration changes. Besides, the employment of batch normalization can also improve the performance. There are respectively 18% and 47% improvement in precision and recall when adding the batch normalization (12 Convolutional layers). This indicates the effectiveness of our applied configurations in OwlEye.

**Table 4: Performance comparison among model configurations (RQ1)**

| Layer number | Without BN | | | With BN | | |
|---|---|---|---|---|---|---|
| | P | R | F1 | P | R | F1 |
| 4 | 0.704 | 0.478 | 0.569 | 0.722 | 0.537 | 0.616 |
| 6 | 0.753 | 0.492 | 0.595 | 0.696 | 0.631 | 0.662 |
| 8 | 0.751 | 0.530 | 0.621 | 0.702 | 0.732 | 0.717 |
| 10 | 0.742 | 0.537 | 0.623 | 0.779 | 0.738 | 0.758 |
| 12 | 0.725 | 0.576 | 0.642 | **0.850** | **0.848** | **0.849** |

*6.1.3 Contribution of Data Augmentation.* We investigate the contribution of data augmentation by comparing the issues detection performance on the overall training data and on the training data by removing the 15,640 screenshots generated with data augmentation (details are in Section 5.2). From Table 5, we can see that both precision (P) and recall (R) improve when the augmented screenshots are added to the training data, indicating the value of data augmentation for effective UI display issues detection. Specifically, 13% and 35% improvement are observed respectively for precision and recall. The larger improvement in recall indicates that, with the augmented dataset, more screenshots with UI display issues can be found. This might because the training set with the argumented data is more diversified in the screenshots, thus has greater issues detection capability.

Table 5: Contribution of data augmentation (RQ1)

| Category | Without DataAug | | | With DataAug | | |
|---|---|---|---|---|---|---|
| | P | R | F1 | P | R | F1 |
| Overall | 0.756 | 0.625 | 0.684 | 0.850 | 0.848 | 0.849 |
| Component occlusion | 0.786 | 0.699 | 0.740 | 0.859 | 0.814 | 0.836 |
| Text overlap | 0.706 | 0.531 | 0.606 | 0.818 | 0.806 | 0.812 |
| Missing image | 0.749 | 0.677 | 0.711 | 0.855 | 0.904 | 0.879 |
| NULL value | 0.742 | 0.356 | 0.481 | 0.855 | 0.808 | 0.830 |
| Blurred screen | 0.857 | 0.600 | 0.705 | 0.888 | 0.800 | 0.842 |

We also present the performance improvement in terms of five categories of UI display issues in Table 5. Results show that, issues detection performance in *null value* category undergoes the largest improvement in F1-score. This might because the training dataset before augmentation has few screenshots (130/3170 = 4%) with this bug (details are in Section 5.2), and the tiny region for deciding this category of bug makes it even difficult for the automatic detection. After adding the augmented screenshots (2188/7820 = 28%), the diversity of the training screenshots significantly improves the performance.

## 6.2 Issues Localization Performance (RQ2)

Figure 8 presents the examples of our issues localization which highlights the buggy areas. We conduct a user study to evaluate the localization performance. We recruit six software developers online, all of whom major in computer science and have more than two years of software development experience. Each of them is presented with 679 correctly detected buggy screenshots in RQ1, and the accompanied localization results as shown in Figure 8.

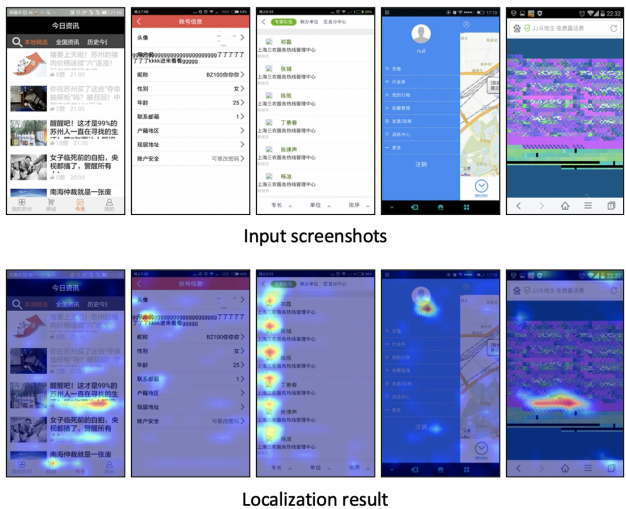

Figure 8: Examples of issues localization (RQ2)

They are required to independently evaluate the issues localization results, and to answer the question whether they agree with each of the localization results using 5-Likert scale, i.e., strongly agree, agree, neutral, disagree, strongly disagree [9, 52]. The evaluation results should be returned within eight hours to ensure the credibility of this study.

As shown in the Table 6, the practitioners strongly agree or agree with the UI display issues localization results in an average of 90% (i.e., 75% + 15%) screenshots, and only disagree (or strongly disagree) in an average of 4% screenshots. This indicates the accuracy of our issues localization in the screenshots. We further calculate the Kendall's W [56] (details are in Section 5.4) to judge to what extent the evaluation results submitted by the six practitioners are consistent with each other. The result of Kendall's W is 0.946, which indicates a high degree of inter-agreement on the performance of our issues localization.

Table 6: Results of issues localization (RQ2)

| Participant | S-Agree | Agree | Neutral | Disagree | S-Disagree |
|---|---|---|---|---|---|
| P1 | 76.1% | 14.3% | 6.2% | 2.7% | 0.7% |
| P2 | 76.6% | 14.1% | 6.0% | 2.4% | 0.9% |
| P3 | 75.1% | 14.9% | 5.7% | 3.0% | 1.3% |
| P4 | 74.3% | 15.5% | 5.9% | 3.1% | 1.2% |
| P5 | 73.8% | 15.2% | 5.7% | 3.2% | 2.1% |
| P6 | 74.4% | 13.6% | 7.2% | 3.2% | 1.6% |
| Average | 75.0% | 14.6% | 6.1% | 2.9% | 1.3% |

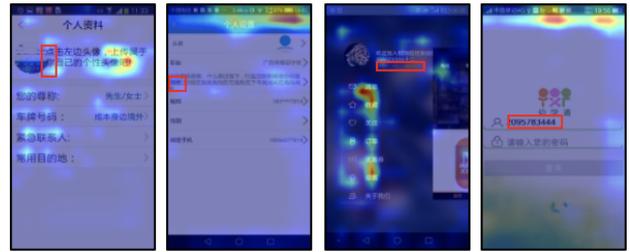

Figure 9: Examples of bad case in issues localization (RQ2)

We further analyze the bad case of issues localization as shown in Figure 9, and find most of them involve the screenshots with *text overlap* and *component occlusion* issues. As mentioned in previous subsection, this might because of the tiny region for localizing the issues which can easily mislead the model.

## 6.3 Usefulness Evaluation (RQ3)

To further assess the usefulness of our OwlEye, we randomly sample 1,500 Android applications from F-droid[7] and 700 Android applications from Google Play[8]. Note that none of these apps appear in our training dataset.

We use DroidBot, which is a commonly-used lightweight Android test input generator [36], for exploring the mobile apps and take the screenshot of each UI pages. Among the 2,200 collected apps, 40% (869/2200) apps can be successfully run with Droidbot, add only 15% (329/2200) of the apps can be fetched with more than one screenshot, as they require register or authenticate to explore more screenshots which cannot be done by DroidBot. For the remaining 329 apps, an average of eight screenshots are obtained for each app. We then feed those screenshots to OwlEye for detecting if there are any UI display issues. Once a display issue is spotted, we create a bug report by describing the issue attached with buggy UI

---
[7]http://f-droid.org/
[8]http://play.google.com/store/apps

screenshot. Finally, we report them to the app development team through issue reports or emails.

Table 7: Confirmed or fixed issues (RQ3)

| APP Name | Category | Source | Download | Id | Status |
|---|---|---|---|---|---|
| Perfect Piano | Music | Google | 50M+ | email | confirm |
| Music Player | Music | Google | 50M+ | email | confirm |
| Nox security | Tool | Google | 10M+ | email | fixed |
| DegooCloud Storage | Tool | Google | 10M+ | email | fixed |
| Proxynel | Tool | Google | 10M+ | email | confirm |
| Secure VPN | Tool | Google | 10M+ | email | confirm |
| Thunder VPN | Tool | Google | 10M+ | email | confirm |
| ApowerMirror | Tool | Google | 5M+ | email | confirm |
| MediaFire | Product | Google | 5M+ | email | confirm |
| Postegro | Commun | Google | 500K+ | email | fixed |
| Deezer Music Player | Music | Google | 500K+ | email | fixed |
| MTG Familiar | Utilities | F-droid | 500K+ | #512 | fixed |
| Open Food Facts | Health | F-droid | 500K+ | #3051 | confirm |
| Linphone | Commun | F-droid | 500K+ | #965 | confirm |
| Paytm | Finance | Google | 100K+ | email | confirm |
| Transdroid | Tool | F-droid | 100K+ | #542 | confirm |
| Transistor | Music | F-droid | 10K+ | #254 | fixed |
| Onkyo | Music | F-droid | 10K+ | #138 | fixed |
| DemocracyDroid | News | F-droid | 10K+ | #51 | confirm |
| NewPipe Legacy | Media | F-droid | 8K+ | #24 | fixed |
| LessPass | Product | F-droid | 5K+ | #519 | fixed |
| CEToolbox | Medical | F-droid | 500+ | #4 | confirm |
| OpenTracks-OSM | Health | F-droid | 10+ | #26 | fixed |
| Yucata Envoy | Tool | F-droid | N/A | #3 | confirm |
| ClassyShark3xodus | Tool | F-droid | N/A | #3 | confirm |
| VlcFreemote | Media | F-droid | N/A | #24 | confirm |

Table 7 shows all bugs spotted by our OwlEye, and more detailed information of detected bugs can be seen in our website[3]. For F-droid applications, 24 UI display issues are detected, among which 6 have been fixed and another 8 have been confirmed by the developers. For Google Play, 33 UI display issues are detected, among which 4 have been fixed and another 8 have been confirmed by the developers. These fixed or confirmed bug reports further demonstrate the effectiveness and usefulness of our proposed approach in detecting UI display issues.

## 7 DISCUSSION

***Generality across platforms.*** Almost all the existing studies of GUI bug detection [32, 36, 67] are designed for a specific platform, e.g., Android, which limits its applicability in real-world practice. In comparison, the primary idea of our proposed OwlEye is to detect UI display issues from the screenshots generated when running the applications with visual understanding. Since the screenshots from different platforms (e.g., Android, iOS) exert almost no difference, our approach can be generalized for UI display issues detection in other platforms. We have conducted a small scale experiment for another popular platform iOS, and experiment on seven screenshots with UI display issues collected in our daily-used applications. Results show that our proposed OwlEye can accurately detect five (71%) of the buggy screenshots. This further demonstrates the generality of OwlEye, and we will conduct more thorough experiment in future.

***Generality across languages.*** Another advantage of our OwlEye is that it can be applied for UI display issues detection in terms of different display languages of the application. The testing data of the experiment for RQ1 contains the screenshots in Chinese, while the experiment for RQ3 relates with the screenshots in English, which demonstrates the generality of our approach across languages. We also collect 22 screenshots with UI display issues in two other languages (i.e. German and Korean) from the applications in RQ3, and run our approach for bug detection. Results show that our proposed OwlEye can accurately detect 16 (73%) of the buggy screenshots, which further demonstrates the feasibility of our OwlEye.

***Potential with more effective automatic testing tool.*** Results in RQ3 have demonstrated the usefulness of OwlEye in real-world practice being integrated with automatic testing tool as DroidBot. However, we have mentioned in Section 6.3 that some applications can not be run with DroidBot, and some can only be fetched with one screenshot due to the shortcoming of DroidBot, both of which limit the full exploration of screenshots. If armed with a more effective automatic testing tool, OwlEye should play a bigger role in detecting UI display issues in real-world practice.

## 8 RELATED WORK

GUI provides a visual bridge between applications and users. Therefore, many researchers are working on assisting developers or designers in the GUI search [5, 10, 12, 16, 28, 53, 69] based on image features, GUI code generation [11, 14, 15, 46, 51] based on computer vision techniques. Moran et al. [47] check if the implemented GUI violates the original UI design by comparing the images similarity with computer vision techniques. A follow-up work by them [50] further detects and summarizes GUI changes in evolving mobile applications. Different from these works, our works are focusing on GUI testing.

To ensure that GUI is working well, there are many static linting tools to flag programming errors, bugs, stylistic errors, and suspicious constructs [13, 70]. For example, Android Lint [1] reports over 260 different types of Android bugs, including correctness, performance, security, usability and accessibility. StyleLint [2] helps developers avoid errors and enforce conventions in styles. Different from static linting, automatic GUI testing [3, 45, 60] dynamically explores GUIs of an app. Several surveys [32, 67] compare different tools for GUI testing for Android apps. Some testing works focus on more specific UI issues such as UI rendering delays [22] and image loading [35]. Recently, deep learning based techniques [18, 66] have been proposed for automatic GUI testing. Unlike traditional GUI testing which explores the GUIs by dynamic program analysis, these two techniques use computer vision techniques to detect GUI components on the screen to determine next actions. Inspired by their works, we also adopt the CNN in our study.

But note that these GUI testing techniques focus on functional testing. In contrast, our work is more about non-functional testing i.e., GUI visual issues which will not cause app crash, but negatively influence the app usability. The UI display bugs detected by our approach are mainly caused by the app compatibility [29, 68] due to the different devices and Android versions. It is highly expensive and extremely difficult for the developers covering all the popular contexts when conducting testing. Besides, different from these works based on static or dynamic code analysis, our work only requires the screenshot as the input. Such characteristic enables our light-weight computer vision based method, and also makes our approach generalised to any platform including Android, IOS, or IoT devices.

## 9 CONCLUSION

Improving the quality of mobile applications, especially in a proactive way, is of great value and always encouraged. This paper focuses on automatic detecting the UI display issues from the screenshots generated during automatic testing. The proposed OwlEye is proven to be effective in real-world practice, i.e., 26 confirmed or fixed previously-undetected UI display issues from popular Android apps.OwlEye also achieves more than 17% and 50% boost in recall and precision compared with the best baseline. As the first work of its kind, we also contribute to a systematical investigation of UI display issues in real-world mobile apps, as well as a large-scale dataset of app UIs with display issues for follow-up studies.

In the future, we will keep improving our model for better performance in the classification. Apart from the display issue detection, we will further locate the root cause of these issues in our future work. Then we will develop a set of tools for recommending patches to developers for fixing display bugs.

## ACKNOWLEDGMENTS

This work is supported by the National Key Research and Development Program of China under grant No.2018YFB1403400.